\documentstyle[twocolumn,aps,epsf,prl]{revtex}
\begin{document}
\flushbottom
\wideabs{
\title{Solitons on the edge of a two-dimensional electron system}
\author{C. Wexler and Alan T. Dorsey}
\address{Department of Physics, University of Florida,
  Gainesville, Florida  32611-8440} 
\date{1 Oct 1998, revised 5 Dec 1998}
\draft
\maketitle

\begin{abstract}
We present a study of the excitations of the edge of a two-dimensional
electron droplet in a magnetic field in terms of a contour dynamics
formalism. We find that, beyond the usual linear approximation, the
non-linear analysis yields soliton solutions which correspond to
uniformly rotating shapes. These modes are found from a perturbative
treatment of a non-linear eigenvalue problem, and as solutions to a
modified  Korteweg-de Vries equation resulting from a local induction 
approximation to the nonlocal contour dynamics. 
We discuss applications to the edge modes in the quantum Hall effect.
\end{abstract}
}   


Shape deformations are important for our understanding of 
such diverse problems as the low-lying excitations of atomic nuclei
\cite{Nuclear}, the hydrodynamics of vortex patches
\cite{ContourAlgorithm,VortexPatchDynamics}, the evolution of
atmospheric plasma clouds \cite{PlasmaClouds}, the formation of 
patterns in magnetic fluids and superconductors
\cite{PatternFormation}, and the dynamics of suspended liquid droplets
\cite{LiquidDropEXPT,LiquidDropMKDV}, to name just a few examples. 
In many of these systems there is a field which is approximately
piecewise constant (for example, the vorticity for  the vortex
patches, or the magnetization for the magnetic fluids); this greatly
simplifies the study of the dynamics, since it is often possible to
focus attention on the boundaries at which this field is
discontinuous, and construct the {\it contour dynamics} for these
boundaries, subject to some global constraint (such as area
conservation).

The edges of a two-dimensional electron system (2DES), and in
particular the edges of a quantum Hall (QH) liquid, present a unique
opportunity to study the dynamics of shape deformations in a clean and
controlled environment. The 2DES in the QH state is incompressible, so
that the electron density is approximately piecewise constant,
suggesting that a contour dynamics approach to studying the droplet
excitations is viable. In addition, the charged nature of the system
facilitates the excitation and detection of deformations of the droplet.

In this paper we will formulate the study of the excitations of a
droplet in a 2DES as a problem in contour dynamics. In the usual
treatment of the edge excitations \cite{EdgesQHE}, a linearization of
the equation of motion is done at early stages, thus limiting the
applicability to small deformations of the edge of the system from an
unperturbed state. In this Letter we consider non-linear terms which
are present in the full contour dynamics treatment. We first present
perturbative results for non-linear deformations of the 2DES shape. We
then show that the local induction approximation to the full contour
dynamics generates the modified Korteweg-de Vries (mKdV) \cite{mKdV}
equation for the curvature dynamics; the mKdV equation also arises in
studies of vortex patches \cite{VortexPatchDynamics} and suspended
liquid droplets \cite{LiquidDropMKDV}.  The mKdV dynamics conserve an
infinite number of quantities, including the area, center of mass, 
and angular momentum of the droplet \cite{LongPaper,KdVHierarchy}, 
so that our local approximation to the nonlocal dynamics preserves the
important conservation laws. The mKdV equation also possesses soliton
solutions, including traveling wave solutions which represent
uniformly rotating deformed droplets.  

{\it I. Hydrodynamics of a two-dimensional electron system 
in a magnetic field.---}Consider a 2DES in a strong magnetic field. The
electron configuration can, in most cases, be characterized by its
density $n({\bf r})$ and its velocity ${\bf v}({\bf r})$. Treated as a
classical fluid, the dynamics is determined by the Euler and
continuity equations,
\begin{eqnarray}
\label{eq:euler1}
\dot{\bf v} =  - \omega_c \; {\bf e}_z &\times& {\bf v} 
- \frac{e^2}{m_e \epsilon} \mbox{\boldmath$\nabla$}
\int d^2r' \; \frac{n({\bf r'})}{|{\bf r}-{\bf r'}|} 
+ \frac{e}{m_e} {\bf E}_{\rm ext}  \;, \\
\label{euler2}
{\partial_t n} &+& \mbox{\boldmath$\nabla$} \cdot ( n
{\bf v}) = 0 \;, 
\end{eqnarray}
where $\omega_c \!=\! e B /m_e$ is the cyclotron frequency and
$\epsilon$ is the dielectric constant of the medium; the first term on
the right hand side of Eq.\ (\ref{eq:euler1}) is the Lorentz force,
the second is the Coulomb interaction, and ${\bf E}_{\rm ext}$ is the
electric field due to the background positive charge, gates, etc. It
is noteworthy that this simple hydrodynamic theory of edge
deformations, and the corresponding canonical quantization of the
classical Lagrangian of the fluid, is also able to capture the essence
of the many-body problem in both integer and fractional QH states
\cite{EdgesQHE}.

The theory of small deformations of the edge has been extensively
studied in Refs.~\cite{EMP,linear,glazman}. The main conclusion is that
for strong magnetic fields, when Landau level quantization
becomes important, the only low energy  modes are edge modes which
propagate in one direction along the edge of the 2DES; the bulk modes
become gaped with a minimum frequency $\omega_c$. For a
sharp electron density profile, this edge mode is the ``conventional''
edge magnetoplasmon mode, with a dispersion relation: 
\begin{equation}
\label{eq:omega0}
\omega_0 (k) = -2 \ln \left( \frac{e^{-\gamma}}{2 |k a|} \right) 
   \frac{\bar{n} e^2}{\epsilon m_e \omega_c} k \;,
\end{equation}
where $k$ is the mode wave-number, $\gamma \!\approx \! 0.5772$ is
the Euler constant, and $a$ is a short-distance cut-off  \cite{EMP}
(the largest of the transverse width of the 2DES, the magnetic length,
or the width of the compressible edge-channel). In addition, for the more
realistic wide and compressible edges, new branches of ``acoustic'' modes 
exist \cite{glazman}. Recent time-of-flight measurements
\cite{AcousticModes} on 2DES in heterostructures confirm this simple
picture. 

\begin{figure}
  \begin{center}
    \leavevmode
    \epsffile{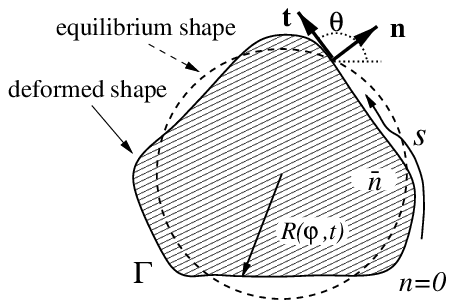}
  \end{center}
  \caption{\label{fig:droplet}
    A charged incompressible liquid in a magnetic field. We assume a
    piecewise constant electron density ($n\!=\! \bar{n}$ inside, while
    $n \!=\! 0$ outside the droplet). The parameterization
    $R(\varphi,t)$, the tangent ${\bf t}$ and normal ${\bf n}$ unit
    vectors to the boundary $\Gamma$ are indicated; $s$ is the
    arclength and $\theta$ the tangent angle. }  
\end{figure}

{\it II. Edge modes: dynamics and kinematics.}---For what follows, 
let us consider the simplest case: a sharp edge between an
incompressible charged fluid and ``vacuum'' (Fig.\
\ref{fig:droplet}). We focus on this mode since it has the simplest
structure and is the most readily observable
\cite{EMP,AcousticModes}. 
As in Ref.\ \onlinecite{EMP}, we neglect
inertial terms on the left hand side of Eq.\ (\ref{eq:euler1}), and
find that the electron velocity is given by 
\begin{equation}
\label{eq:vbounbd1}
{\bf v({\bf r})} = - \frac{e^2}{\epsilon m_e \omega_c}
\mbox{\boldmath$\nabla$} \times {\bf e}_z 
        \int_{\cal A} d^2r' \; \frac{ n({\bf r'})}
	{|{\bf r}-{\bf r'}|} + 
        {\bf v}_{\rm ext} \;,
\end{equation}
where ${\bf v}_{\rm ext} \!=\! - (e/m_e \omega_c) \: {\bf e}_z \times
{\bf E}_{\rm ext}$ is the velocity induced by the external field, and 
${\cal A}$ is the area of the droplet. 

Let us now concentrate on the ``internal'' velocity given by the first
term in Eq.\ (\ref{eq:vbounbd1}) \cite{external}. For an
incompressible 2DES with a piecewise constant density
\cite{compressible}, the density can
be taken outside the integral; then using Stokes' theorem, the area
integral can be transformed into a line integral over the boundary
$\Gamma \!=\! \partial {\cal A}$ of the electron liquid:
\begin{equation}
\label{eq:vlineint}
{\bf v({\bf r})} = \frac{\bar{n} e^2}{\epsilon m_e \omega_c} 
        \oint_\Gamma ds' \,\frac{ {\bf t}(s')}{|{\bf r}-{\bf r}(s')|} \;.
\end{equation}
Here ${\bf t}(s')\!=\!\partial_{s'}{\bf r}(s') \!\equiv\! {\bf r}_{s'}$ 
is the unit tangent vector at the arc-length $s'$; for later use, we
also define ${\bf n}\!\perp\!{\bf t}$ as the unit normal vector. The
short distance singularity in the integrand is cut off at a length scale
$r_0$. Equation (\ref{eq:vlineint}) forms the basis of our contour
dynamics treatment---it expresses the velocity of the edge in terms of
a  nonlocal self-interaction of the edge. 

For the sake of comparison, we draw analogy to the case of a vortex 
patch, a two-dimensional, bounded region of constant vorticity
$\omega_p$ surrounded by an irrotational fluid, where the interaction
is logarithmic \cite{ContourAlgorithm,VortexPatchDynamics}: 
\begin{equation}
\label{eq:vp}
{\bf v({\bf r})} = - \frac{\omega_p}{2 \pi} 
        \oint_\Gamma ds' \, {\bf t}(s') \,
        \ln \left[{|{\bf r}-{\bf r}(s')|\over r_0} \right]\;.
\end{equation}
The vorticity and the area of the patch are conserved, a consequence
of Kelvin's circulation theorem \cite{batchelor}. We see that in both
cases (i) the dynamics is chiral, being determined by the tangent
vector; (ii) the fluid contained within $\Gamma$ is incompressible, so
that the area  is conserved.  It is this analogy which inspired the
present work. 

Having determined the velocity of the electron liquid, we now focus on
the motion of the 2DES boundary $\Gamma$. The velocity of a point on the
boundary can be written in terms of the normal and tangential components, 
\begin{equation}
{\bf v} = U(s)\, {\bf n} + W(s)\, {\bf t}. 
\end{equation}
The tangential velocity $W(s)$ is largely irrelevant. 
We now ask whether there are modes which propagate
along the boundary with no change in shape. Previous work \cite{EMP}
has focused on small perturbations of a straight, infinite edge. Here
we consider deformations of a circular droplet of incompressible
electrons. A uniformly propagating mode is therefore characterized by
a boundary that moves like a rotating rigid body,
namely the radius of the boundary satisfies $R(\varphi,t) \!=\!
R(\varphi \!-\! \Omega t)$, where $\varphi$ is the azimuthal angle and
$\Omega$ is the angular frequency.
This translates into a condition for the 
{\em normal} velocity:
\begin{equation}
\label{eq:rigidrotation}
U \equiv \left. {\bf n}({\bf r})\cdot{\bf v}({\bf r}) 
        \right|_{{\bf r} \in \Gamma} = 
\Omega \; {\bf n}({\bf r})\cdot \left( {\bf e}_z \times {\bf r}\right)
\;.
\end{equation}

We seek boundary shapes $R(\varphi)$ (see Fig.\ \ref{fig:droplet})
that rotate uniformly, satisfying Eq.\ (\ref{eq:rigidrotation}). 
Consider the following parameterization of the surface:   
\begin{equation}
\label{eq:paramet}
R(\varphi) = R_0 \left( 1 + \sum_{l=-\infty}^{\infty} b_l \; 
        e^{i l \varphi} \right) \;.
\end{equation}
The unit tangent vector is given by ${\bf t}(\varphi) \!=\!
\mbox{\boldmath$\tau$}(\varphi)/ |\mbox{\boldmath$\tau$}(\varphi)|$
where $\mbox{\boldmath$\tau$}(\varphi) \!\equiv \! \partial {\bf
  r(\varphi)} / \partial \varphi \!=\! {\bf e}_r R'(\varphi) \!+\!
{\bf e_\varphi} R(\varphi)$. Likewise, the unit normal vector is given
by ${\bf n}(\varphi) \!=\! - {\bf e}_z \times 
\mbox{\boldmath$\tau$}(\varphi)/ |\mbox{\boldmath$\tau$}(\varphi)|$. 
The normal velocity on the boundary can be written as
\begin{equation}
\label{eq:normalv}
U (\varphi) = \frac{\bar{n} e^2}{\epsilon m_e \omega_c} 
        \int_0^{2 \pi} \! d\varphi' \; 
        \frac{{\bf n}(\varphi) \cdot \mbox{\boldmath$\tau$}(\varphi') }
                {|{\bf R}(\varphi) - {\bf R}(\varphi')|} \;.
\end{equation}

It has not been possible to solve exactly the {\em non-linear
eigenvalue problem} for $b_l$ and $\Omega$ [Eqs.\
(\ref{eq:rigidrotation}-\ref{eq:normalv})] exactly. We therefore seek
a perturbative solution by expanding the right-hand side of Eq.\
(\ref{eq:normalv}) in powers of $b_l$. This allows us to to go beyond
the linear approximations used in the past, and we have succeeded in
calculating shape deformations to ${\cal O}[b_l^4]$ and angular
frequencies to ${\cal O}[b_l^5]$. Expanding the non-linear eigenvalue
problem to fifth order, we find the condition
\begin{eqnarray}
\label{eq:NLEP}
\tilde{\Omega} &&\left(b_l + \frac{1}{2} 
\sum_p b_{l-p} b_p \right) =
Q_l\, b_l 
+ \sum_p R_{l,p} \, b_{l-p} b_p \nonumber \\
&&+\! \sum_{p,q} S_{l,p,q} \, b_{l-p} b_{p-q} b_q  
+ \!\sum_{p,q,r} T_{l,p,q,r} \, b_{l-p} 
        b_{p-q} b_{q-r} b_r \nonumber \\
&&+\!\!\! \sum_{p,q,r,s} U_{l,p,q,r} 
        b_{l-p} b_{p-q} b_{q-r} b_{r-s}
b_s + {\cal O} [b_l^6] \;, 
\end{eqnarray}
where $\tilde{\Omega} \!=\! ({\epsilon m_e \omega_c R_0}/{\bar{n}
e^2})\, \Omega$, and the ``matrix elements'' $Q$, $R$, $S$, $T$ and $U$ are
obtained from Eq.\ (\ref{eq:normalv}). The details of this calculation
will be published elsewhere \cite{LongPaper}. The zeroth
order angular frequency for the eigenmode with rotational
symmetry $C_l$ (rotations by $2\pi/l$) is
\begin{equation}
\label{eq:Q}
\tilde{\Omega}^{(0)}[l]  =  Q_l =
4  \sum_{k=2}^{|l|}\frac{1}{2k-1},
\end{equation}
where the last sum can be related to the digamma function 
$\psi(|l|\!+\!{\small 1/2})$. 
This linear result has been
previously derived by several authors \cite{linear}; corrections are
${\cal O}[b_l^2]$. For a direct comparison with Eq.\
(\ref{eq:omega0}), which corresponds to the {\em large-$l$} limit, we
substitute the asymptotic expansion for the sum in Eq.\ (\ref{eq:Q});
multiplication by $R_0$ yields the propagation velocity: 
\begin{equation}
v_g = -2 \ln \left( \frac{e^{-\gamma}}{4 \, e^2 |l|} \right)
        \frac{\bar{n} e^2}{\epsilon m_e \omega_c} \;,
\end{equation}
which closely corresponds to $v_g \!\equiv\! \partial \omega_0(k) / 
\partial k$ after the substitution $l\!\sim\!ka$ [see Eq.\ 
(\ref{eq:omega0})]. The dispersion for these linear edge excitations
has been confirmed experimentally in both the frequency
\cite{FreqDomain} and time \cite{AcousticModes,TimeDomain} domains.

\begin{figure}
  \begin{center}
    \leavevmode
    \epsffile{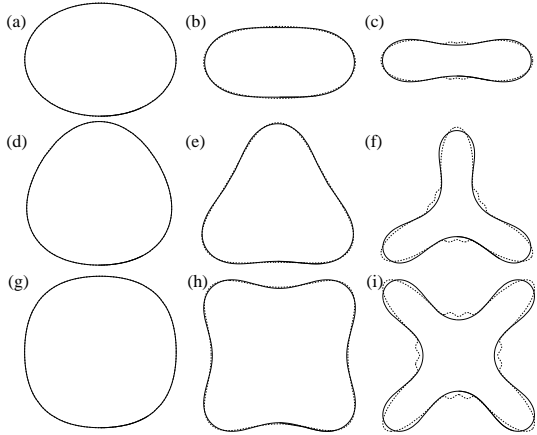}
  \end{center}
  \caption{ \label{fig:shapes} 
        Uniformly rotating shapes of a 2DES. Solid lines: solutions of
the mKdV equation obtained from the local induction
approximation. Dotted lines: solutions obtained using the perturbative
expansion. The values of the coefficient $b_l$, and the ratio of
curvatures $\sigma\!\equiv\!\kappa_{\rm min}/\kappa_{\rm max}$ are:
{(a)} $b_2 \!=\! 0.073, \, \sigma \!=\! 0.4$;
{(b)} $b_2 \!=\! 0.19, \, \sigma \!=\! 0$;
{(c)} $b_2 \!=\! 0.36, \, \sigma \!=\! -\!0.2$;
{(d)} $b_3 \!=\! 0.027, \, \sigma \!=\! 0.4$;
{(e)} $b_3 \!=\! 0.10, \, \sigma \!=\! -\!0.2$;
{(f)} $b_3 \!=\! 0.29, \, \sigma \!=\! -\!0.45$;
{(g)} $b_4 \!=\! 0.014, \, \sigma \!=\! 0.4$;
{(h)} $b_4 \!=\! 0.089, \, \sigma \!=\! -\!0.4$;
{(i)} $b_4 \!=\! 0.24, \, \sigma \!=\! -\!0.56$. }
\end{figure}

Some of the shapes are shown by the dotted curves in Fig.\
\ref{fig:shapes}, and closely resemble the ``V states'' of vortex
patches found by Deem and Zabusky \cite{VStates}. For larger
deformations, the appearance of oscillations indicate that higher
order terms are needed, since they correspond to higher Fourier
components. 

{\it III. Local induction approximation.}---As we have seen, 
the motion of the edge is determined by the velocity of the fluid at
the surface. The nonlocal equation for the velocity of the boundary,
Eq.\ (\ref{eq:vlineint}), can be turned into a differential equation
for the {\em curvature} of the boundary if we concentrate on the local
contributions. This {\em local induction approximation} (LIA)
\cite{batchelor} was explored by Goldstein and Petrich in a 
series of papers dealing with the evolution of vortex patches
\cite{VortexPatchDynamics,KdVHierarchy}. The situation is
considerably more favorable in this problem, due to the more rapid 
decay of the  interaction [$1/r$ for charges vs.\ $\ln(r)$ for
vortices, see Eqs.\ (\ref{eq:vlineint}) and (\ref{eq:vp})]. 

The LIA is an expansion of the integrand of Eq.\ (\ref{eq:vlineint})
about $s' \!=\! s$; a long-distance cut-off $\Lambda$ is introduced by
replacing the line integral  $\oint_\Gamma \{\cdots\} ds'$ by  
$\int_{s-\Lambda/2}^{s+\Lambda/2} \{\cdots\} ds'$. Using the
Frenet-Serret relations ${\bf r}_s \!=\! {\bf t}$,  
${\bf r}_{ss} \!=\! {\bf t}_s \!=\! -\kappa \, {\bf n} $,
where $\kappa\!=\!\theta_s$ is the local curvature of the boundary, to
lowest order the normal and tangential velocities are \cite{LongPaper}: 
\begin{eqnarray} 
\label{eq:LIAresults}
&&U_{\rm LIA} = - \left[\frac{\bar{n} e^2}{\epsilon m_e \omega_c}\right]
         \frac{\Lambda^2}{8} \, \kappa_s \;, \nonumber \\
&&W_{\rm LIA} = \left[\frac{\bar{n} e^2}{\epsilon m_e \omega_c}\right]
         \left( \ln \frac{\Lambda^2}{2 r_0} - \frac{11 \Lambda^2}{96}
           \, \kappa^2 \right) \;.
\end{eqnarray}
Since the rate of change of the area of the droplet is 
${\cal A}_t \!=\! \oint ds\, U(s)$, the LIA, with 
$U_{\rm LIA} \!\propto\! \kappa_s$, automatically conserves area. 

The time evolution of a curve in two dimensions is given quite
generally by the differential equation \cite{KdVHierarchy,Brower}
\begin{equation}
\label{eq:diffKappa}
\kappa_t = - [ \kappa^2+\partial_{ss} ] U + \kappa_s W - \kappa_s
\int_0^s [\kappa U + W_{s'}]ds' \;.
\end{equation}
Introducing the results from Eq.\ (\ref{eq:LIAresults}), changing 
``gauge'' (by modifying $W$) and rescaling time, we find that the
curvature satisfies the mKdV equation \cite{mKdV}: 
\begin{equation}
\label{eq:mkdv}
\kappa_t = 
\frac{3}{2} \kappa^2 \kappa_s +
  \kappa_{sss}  \;.
\end{equation}
The mKdV dynamics are integrable, with an infinite number of globally
conserved geometric quantities \cite{VortexPatchDynamics}, the most
important of which are the center of mass, area, and angular
momentum of the droplet.

The mKdV equation possesses a variety of soliton solutions, 
including traveling wave solutions and propagating ``breather''
solitons. Here we will focus on the traveling wave solutions of Eq.\
(\ref{eq:mkdv}) of the form $\kappa(s,t)\!=\!g(z)$ with
$z\!\equiv\!s \!-\!ct$, which represent uniformly rotating deformed
droplets. The ordinary differential equation for $g(z)$ can be
integrated twice with the result 
\begin{equation}
\label{eq:ode}
\frac{1}{2} (g')^2 = -\frac{1}{8} g^4 +\frac{1}{2} c g^2+ a g - 2 b
\;,
\end{equation}
where $a$ and $b$ are constants of integration ($a\!=\!b\!=\!0$ for
infinite systems). The periodic solutions of this equation, expressed
in terms of Jacobi elliptic functions, 
are:
\begin{eqnarray}
\label{eq:curvatureMKDV}
\kappa(z) &=& \frac{(q \kappa_{\rm max} \!+\! p \kappa_{\rm min}) \!+\!
  (p \kappa_{\rm min} \!-\! q \kappa_{\rm max})\,
  {\rm cn}\! \left( \!\sqrt{pq} \frac{\displaystyle z}{ 2}\,
        |\, \lambda \right) }
{(p+q)+(p-q)\,{\rm cn}\! \left( \!\sqrt{pq} 
\frac{\displaystyle z}{ 2}\,|\,\lambda \right)}
\,, \nonumber \\
p^2 &=& {(3 \kappa_{\rm max} + \kappa_{\rm min})^2 +
\xi^2} \,,\nonumber \\ 
q^2 &=&  {(\kappa_{\rm max} + 3\kappa_{\rm min})^2 +
\xi^2} \,,\nonumber \\ 
\lambda^2 &=& { \left[(\kappa_{\rm max} - \kappa_{\rm min})^2 -
    (p-q)^2\right]/4 p q} \,,
\end{eqnarray}
where $\kappa_{\rm max, min}$ are the maximum and minimum curvatures,
and $\xi$ is determined by the boundary conditions. 
The period of $\kappa$ is given by the elliptic integral
$L_\kappa \!\equiv\! (8/\!\sqrt{p q}) \,K \!(\lambda)$. The first
integral of the curvature gives the tangent angle $\theta(s) \!=\!
\int_0^s \kappa(s')ds'$. We require that $\theta(l\,L_\kappa) \!=\! 2
\pi$, so that the resulting curve is closed (this fixes $\xi$). The
integer $l$ determines that the curve has $C_l$ symmetry.
The curves thus generated can be characterized by
$(l,\kappa_{\rm max},\kappa_{\rm min})$ or more conveniently by the
symmetry, the area and the perimeter of the curve. 

The full lines in Fig.\ \ref{fig:shapes} show some uniformly rotating
soliton shapes,  calculated from Eq.\ (\ref{eq:curvatureMKDV}). These
are essentially identical  with Goldstein and Petrich's
\cite{KdVHierarchy} soliton solutions for the vortex patch problem (Deem
and Zabusky's ``V states'' \cite{VStates}), but the more local
interaction in the QH case guarantees a better correspondence with the
perturbative solutions. Indeed, the curves resulting from the
perturbative method and the LIA are quite close, even for considerable
deformations of the boundary. For larger deformations the perturbative
results show artifacts due to the limited number of Fourier
components. The advantage of the LIA becomes evident in this case, 
since it is an expansion in powers of the {\em curvature} and not the
{\it deformation}, and thus allows for relatively large 
long-wavelength deformations. More significantly, the LIA and the
resulting integrable dynamics allow one to uncover geometrical
conservation laws which would be hidden in a perturbative calculation
\cite{LongPaper}. This advantage comes at a price: the detailed 
information on frequencies is obscured by the introduction of the long
distance cut-off $\Lambda$ and by the gauge transformation of the
tangential velocity $W$, while the frequency is easily obtained in the
perturbative calculation. 

{\it IV. Conclusions.}---A contour dynamics formulation of the
excitations on the edge of a two-dimensional electron system in a
magnetic field has allowed us to demonstrate the existence, beyond
the usual linear regime, of shape deformations that propagate
uniformly.  A local approximation to the nonlocal dynamics shows that 
the curvature of the edge of the droplet obeys the modified 
Korteweg-de Vries equation, which has integrable dynamics and soliton
solutions. Since these solutions are dispersionless, it may be
possible to distinguish them from linear edge waves in time-of-flight
measurements.  Earlier studies \cite{zhitenev} have shown non-linear
waves in edge channels, but the origin of the non-linearity and
experimental details are different than considered here.
On the theoretical side, it would be interesting to
connect our hydrodynamic treatment of these edge solitons with 
field-theoretical treatments of edge excitations \cite{linear}.


We would like to thank Raymond Goldstein for useful discussions.
This work was supported by the NSF grant DMR-9628926.


\end{document}